%Paper: hep-ph/9311300
%From: JPHALYO@wiswic.weizmann.ac.il
%Date: Tue, 16 Nov 1993 13:51:15 GMT

% Printing instructions:
%       This paper needs the macro packages phyzzx.tex and tables.tex
%       The paper contains two tables, appended at the end of the TeX
%       file. The tables should be stripped off and printed separately.
%
\input phyzzx
\tolerance=1000
\sequentialequations
\def\rl{\rightline}

\def\r#1{$\bf#1$}

\def\t1{{\tilde 1}}

\def\AEF{A.E. Faraggi}
\def\DVN{D. V. Nanopoulos}

\def\NPB#1#2#3{Nucl. Phys. B {\bf#1} (19#2) #3}
\def\PLB#1#2#3{Phys. Lett. B {\bf#1} (19#2) #3}

\def\PRL#1#2#3{Phys. Rev. Lett. {\bf#1} (19#2) #3}
\def\PRT#1#2#3{Phys. Rep. {\bf#1} (19#2) #3}

\def\IJMP#1#2#3{Int. J. Mod. Phys. A {\bf#1} (19#2) #3}

\def\l{\langle}
\def\r{\rangle}

\REF\THOOFT{G. 't Hooft, in Recent Developements in Gauge Theories, ed. G. 't
Hooft et. al. (Plenum, New York, 1980).}
\REF\SCP{J.E. Kim, \PRT{150}{87}{1}; H.Y. Cheng, \PRT{158}{88}{1}.}
\REF\GM{H. Georgi and I. N. McArthur, Harvard University Report HUTP--81/A001
(1981), unpublished: D. Kaplan and A. Manohar, \PRL{56}{86}{2004}.}
\REF\GSW{M. Green, J. Schwarz and E. Witten,
Superstring Theory, 2 vols., Cambridge
University Press, 1987.}
\REF\MOD{\AEF, \PLB{278}{92}{131}.}
\REF\SLM{\AEF, \PLB{274}{92}{47}; \NPB{387}{92}{289}.}
\REF\DSW{M. Dine, N. Seiberg and E. Witten, \NPB{289}{87}{585}.}
\REF\ADS{J.J. Atick, L.J. Dixon and A. Sen, \NPB{292}{87}{109};
         S. Cecotti, S. Ferrara and M. Villasante, \IJMP{2}{87}{1839}.}
\REF\KLN{S. Kalara, J. Lopez and D.V. Nanopoulos, \PLB{245}{91}{421};
\NPB{353}{91}{650}.}
\REF\CKM{A. E. Faraggi and E. Halyo, \PLB{307}{93}{305}; WIS--93/34/MAR--PH.}
\REF\FM{\AEF, WIS--92/81/OCT--PH, to appear in Nucl. Phys. B.}
\REF\FFF{I. Antoniadis, C. Bachas, and C. Kounnas, \NPB{289}{87}{87};
I. Antoniadis and C. Bachas, \NPB{298}{88}{586};
5H. Kawai, D.C. Lewellen, and S.H.-H. Tye,
Phys. Rev. Lett. {\bf57} (1986) 1832;
Phys. Rev. D {\bf 34} (1986) 3794;
Nucl. Phys. B {\bf 288} (1987) 1;
R. Bluhm, L. Dolan, and P. Goddard,
Nucl. Phys. B {\bf 309} (1988) 330.}
\REF\REVAMP{I. Antoniadis, J. Ellis, J. Hagelin, and \DVN, \PLB{231}{89}{65}.}
\REF\NRT{\AEF, \NPB{403}{93}{101}.}
%\REF\GCU{A.E. Faraggi, Phys. Lett. B {\bf302} (1993) 202.}
\REF\AX{E. Halyo, WIS--93/66/JUL--PH, to appear in Physics Letters B.}

\singlespace
\rl{WIS--93/98/OCT--PH}
\rl{\today}
\rl{T}
\pagenumber=1
\normalspace
\smallskip
\titlestyle{\bf{Supersymmetry and Light Quark Masses in a Realistic
Superstring Model}}
\smallskip
\author{Edi Halyo{\footnote\dag{e--mail address: jphalyo@weizmann.bitnet}}}
\smallskip
\centerline {Department of Physics, Weizmann Institute of Science}
\centerline {Rehovot 76100, Israel}
\vskip 6 cm
\titlestyle{\bf ABSTRACT}

We examine the light quark masses in a standard--like superstring model in
the four dimensional free fermionic formulation. We find that the supersymmetry
constraints in the observable and hidden sectors eliminate all large
contributions to $m_u$ and $m_d$ and force them to be much smaller than the
other quark masses. The requirement for an acceptable Higgs doublet spectrum
results in $m_u<<m_d$. In these models
a realistic $m_d$ can always be obtained whereas $m_u$ is at most
$10^{-5}~MeV$.
For particular choices of flat directions or vacua $m_u$ can be as small as
$10^{-7}~MeV$ but cannot vanish.

\singlespace
\vskip 0.5cm
\endpage
\normalspace

\centerline{\bf 1. Introduction}

One of the many puzzling features of the quark spectrum is the smallness of
the (current) up and down quark masses. These are not only suppressed by a
factor of $\sim 10^{-4}-10^{-5}$ with respect to the weak scale but are also
much smaller than the other quark masses.
According to 't Hooft's naturalness criterion [\THOOFT], the smallness of
$m_u$ and $m_d$ must follow from symmetries which are
broken only by a very small amount in order to result in such small
$m_u$ and $m_d$. In the limit of exact symmetry one would expect to have
vanishing $m_u$ and $m_d$. Thus, a small or vanishing $m_u$ is only
natural if it is a result of some (possibly discrete) symmetry.
Another aspect of a small or vanishing $m_u$ is that $m_u=0$ is a possible
solution to the
strong CP problem [\SCP]. By now, it is well known that a vanishing $m_u$ is
not in conflict with current algebra results [\GM].

Any extension of the standard model which tries to explain the origin of
fermion
masses must explain or at least accomodate the light fermion masses. Foremost
among these are superstring theories [\GSW]. Certainly, if superstring theories
are ``the theories of everything", they should explain the smallness of $m_u$
and $m_d$ in addition to the rest of the quark spectrum. It is therefore
important to examine light quark masses in realistic superstring models. The
purpose of this work is to
see whether current up and down quark masses can be obtained in standard--
like superstring models. Moreover, we would like to know if a vanishing $m_u$
is possible. We also hope to gain an understanding of the symmetries which
cause the light quark masses to be much smaller than those of other quarks.

The standard--like superstring model that we consider has the following
properties [\MOD,\SLM]:

1. $N=1$ space--time supersymmetry (SUSY).

2. A $SU(3)_C\times SU(2)_L\times {U(1)^n}\times$hidden gauge group.

3. Three generations of chiral fermions
and their superpartners, with the correct quantum numbers
under ${SU(3)_C\times SU(2)_L\times U(1)_Y}$.

4. Higgs doublets that can  produce realistic electro--weak symmetry breaking.

5. Anomaly cancellation, apart from a single ``anomalous" U(1)
which is  canceled by  application of the
Dine--Seiberg--Witten (DSW) mechanism [\DSW].

As noted above, in standard--like superstring models there are no
gauge and gravitational anomalies apart from a single ``anomalous
$U(1)$" symmetry. This anomalous $U(1)_A$ generates a Fayet--Iliopoulos term
that breaks SUSY at the Planck scale [\DSW]. SUSY is restored
and $U(1)_A$ is broken by giving VEVs to a set of standard model singlets in
the
massless string spectrum along the flat F and D directions [\ADS].
Thus, the $SO(10)$ singlet fields in the non--renormalizable terms
obtain non--vanishing VEVs by the application of the DSW mechanism.
In addition, scalars which are in the vector representations of unbroken,
non--Abelian, hidden sector gauge groups condense and obtain VEVs at the
condensation scale, $\Lambda_H$. Then, the order $N(=m+n)$
non--renormalizable terms, of the form $cffh(\Phi^m V^n/M)^{N-3}$ (for
$n=0,2$),
become effective Yukawa terms, where $f,h,\Phi,V$ denote fermions, scalar
doublets, scalar singlets and hidden sector states, respectively. $M$ is a
Planck scale mass to be defined later. The effective Yukawa couplings are
given by $\lambda=c(\langle \Phi^m V^n \rangle/M)^{N-3}$
where the calculable coefficients $c$ are of order one [\KLN].
In this manner quark mass terms, as well as quark mixing terms, can be
obtained. Realistic quark mixing and masses for the two heavy generations have
been obtained for a suitable choice of scalar VEVs [\CKM].

In Ref. (11)  the quark mass hierarchy for the heaviest two generations was
obtained by giving mass from the cubic superpotential only to the top quark.
The other quarks except the light ones (i.e. $u$ and $d$) get their masses from
$N=5$ non--renormalizable terms and hence they are suppressed relative to
the top mass by a factor of $\l \Phi^2 \r/M^2 \sim 10^{-2}-10^{-3}$. It was
also noticed that light quark masses cannot arise only from the observable
sector VEVs, to any order in $N$, due to SUSY constraints in the observable
sector.

In this work we examine the contributions to light
quark masses that arise from hidden sector VEVs in addition to those from the
observable sector. We find that SUSY constraints in the observable and hidden
sectors (with the requirement of realistic heavy quark masses) eliminate the
potentially large contributions to up and down masses. In fact, SUSY constrains
$m_u$ and $m_d$ to be around the $MeV$ scale.  This can be explained
as a result of an effective $Z_4$ symmetry arising from the SUSY constraints.
There are two possible scenarios in which either $\bar h_1$ or $\bar h_2$
is the light Higgs doublet that couples to the up--like quarks. In both cases
the doublet that couples to down--like quarks is $h_{45}$. We find that a
realistic $m_d \sim$ $MeV$ can be obtained in both cases from the
off--diagonal terms in the down quark mass matrix. On the other hand, we find
that $m_u$ is at most $\sim 10^{-5}~MeV$ which is six orders of magnitude
smaller than its value obtained from current algebra i.e. $\sim 5~MeV$ but not
small enough to solve the strong CP problem naturally.
If we take $\l \bar \Phi_3^+ \r=0$, then the only contribution to $m_u$ comes
from $N=8$ terms which contain hidden sector condensates and give $m_u \sim
10^{-6}~MeV$. We also find that $m_u$ vanishes up to $N=8$ for a flat
direction with $\l \Phi_3^\pm \r= \l \bar \Phi_3^\pm \r=0$. There are always,
non--zero, order $N>8$ terms
which contribute to $m_u$ and therefore $m_u$ cannot vanish to all orders in
$N$. $N=9$ terms give a lower bound of $m_u \sim 10^{-7}~MeV$.
$m_u<<m_d$ in this model since the only $MeV$ scale contribution to $m_u$
vanishes due to the constraints from an acceptable Higgs doublet spectrum as
we show below.

The paper is organized as follows. In Section 2, we review the
superstring model. In Section 3, we find the SUSY constraints on VEVs of the
hidden sector states in addition to those of the observable sector states.
In Section 4, we examine the Higgs doublet mass matrix and find the
constraints for a realistic Higgs doublet spectrum. In Section 5,
we obtain the up and down quark mass matrices and examine the contributions to
the light quark masses in detail. We also consider all possible kinds of
non--renormalizable terms which contribute to $m_u$ and $m_d$.
In Section 6, we present a discussion and our conclusions.

\bigskip
\centerline{\bf 2. The superstring model}

The superstring standard--like models are constructed in the four
dimensional free fermionic formulation [\FFF].
The models are generated by a basis of eight boundary condition vectors
for all world--sheet fermions. The first five vectors in the basis
consist of the NAHE
set $\{{\bf 1},S,b_1,b_2,b_3\}$ [\REVAMP].
The standard--like models are constructed by adding three additional
vectors to the NAHE set [\MOD,\SLM].
The observable and hidden gauge groups after application
of the generalized GSO projections are $SU(3)_C\times U(1)_C\times
 SU(2)_L\times U(1)_L\times U(1)^6${\footnote*{$U(1)_C={3\over 2}U(1)_{B-L}$
and $U(1)_L=2U(1)_{T_{3_R}}$.}}
and $SU(5)_H\times SU(3)_H\times U(1)^2$, respectively.
The weak hypercharge is given by
$U(1)_Y={1\over 3}U(1)_C + {1\over 2}U(1)_L$ and has the standard $SO(10)$
embedding. The orthogonal
combination is given by $U(1)_{Z^\prime}= U(1)_C - U(1)_L$.
The model has six right--handed and six left--handed horizontal symmetries
$U(1)_{r_j}\times U(1)_{\ell_j}$ ($j=1,\ldots,6$),
which correspond to the right--moving and left--moving world--sheet
currents respectively.

The full massless spectrum with the quantum numbers was presented in Ref.
[\MOD].
Here we list only the states that are relevant for the quark mass matrices.

(a) The $b_{1,2,3}$ sectors produce three $SO(10)$ chiral generations,
$G_\alpha=e_{L_\alpha}^c+u_{L_\alpha}^c+N_{L_\alpha}^c+d_{L_\alpha}^c+
Q_\alpha+L_\alpha$ $(\alpha=1,\cdots,3)$

(b) The ${S+b_1+b_2+\alpha+\beta}$ sector gives the weak doublet $h_{45}$, the
color triplet $D_{45}$ and the $SO(10)$ singlets $\Phi_{45},\Phi_1^\pm,
\Phi_2^\pm,\Phi_3^\pm$ and their conjugates.

(c) The Neveu--Schwarz $O$ sector gives, in addition to  the graviton,
dilaton, antisymmetric tensor and spin 1 gauge bosons, the scalar
weak doublets $h_1,h_2,h_3$ and the singlets $\Phi_{12},\Phi_{23},
\Phi_{13}$ (and their conjugates).
Finally, the Neveu--Schwarz sector gives rise to three singlet
states $\xi_{1,2,3}$ that are neutral under all the U(1) symmetries.

The sectors $b_j+2\gamma+(I){\hskip .2cm} (j=1,..,3)$ give vector--like
representations that are
$SU(3)_C\times SU(2)_L\times {U(1)_L}\times {U(1)_C}$
singlets and transform as $5$, ${\bar 5}$ and $3$, ${\bar 3}$
under the
hidden $SU(5)$ and $SU(3)$ gauge groups, respectively (see Table 1).
The states from the sectors $b_j+2\gamma+(I)$ produce the mixing between the
chiral generations [\CKM]. In addition, they give masses to the light Higgs
doublets and light quarks from non--renormalizable terms in the superpotential
as we will see below.

The massless spectrum also contains states from sectors with some combination
of $\{b_1,b_2,b_3,\alpha,\beta\}$ and $\gamma+(I)$. These states are
model dependent and carry either fractional electric charge or
$U(1)_{Z^\prime}$ charge. The ones with fractional charges get large masses
from non--renormalizable interactions and decouple from the spectrum. The
states with $U(1)_{Z^\prime}$ charge are listed in Table 2. These appear in
the trilevel superpotential and mix with states from the observable sector.
As a result, we will see that there are non-trivial constraints on the VEVs
of these from SUSY and an aceptable Higgs doublet spectrum.

%The structure of the massless spectrum
%exhibited in Eqs. (1-4), and in Tables 1 and 2, is common to a large number
%of free fermionic standard--like models. All the standard--like models
%contain three chiral generations from the sectors $b_j$, vector--like
%representations from the sectors $b_j+2\gamma$, Higgs doublets from
%the Neveu--Schwarz sector and hidden sector states with non--zero $U(1)_{Z^
%\prime}$ charge.

In addition to the spectrum, we have to consider the superpotential of the
model. Cubic and non--renormalizable
contributions to the superpotential are obtained by calculating correlators
between vertex operators [\KLN]
$A_N\sim\langle V_1^fV_2^fV_3^b\cdot\cdot\cdot V_N^b\rangle $
where $V_i^f$ $(V_i^b)$ are the fermionic (bosonic) vertex operators
corresponding to different fields. The non--vanishing terms are obtained by
applying the rules of Ref. [\KLN].

At the cubic level the following terms are obtained in the observable and
hidden
sectors [\MOD],
$$\eqalignno{W_3&=\{(
{u_{L_1}^c}Q_1{\bar h}_1+{N_{L_1}^c}L_1{\bar h}_1+
{u_{L_2}^c}Q_2{\bar h}_2+{N_{L_2}^c}L_2{\bar h}_2+
{u_{L_3}^c}Q_3{\bar h}_3+{N_{L_3}^c}L_3{\bar h}_3)\cr
&+{{h_1}{\bar h}_2{\bar\Phi}_{12}}+{h_1}{\bar h}_3{\bar\Phi}_{13}
+{h_2}{\bar h}_3{\bar\Phi}_{23}+{\bar h}_1{h_2}{\Phi_{12}}
+{\bar h}_1{h_3}{\Phi_{13}}+{\bar h}_2{h_3}{\Phi_{23}}
+\Phi_{23}{\bar\Phi}_{13}{\Phi}_{12}\cr
&+{\bar\Phi}_{23}{\Phi}_{13}{\bar\Phi}_{12}
+{\bar\Phi}_{12}({\bar\Phi}_1^+{\bar\Phi}_1^- +{\bar\Phi}_2^+{\bar\Phi}_2^-
+{\bar\Phi}_3^+{\bar\Phi}_3^-)+{\Phi_{12}}(\Phi_1^-\Phi_1^+
+\Phi_2^-\Phi_2^++\Phi_3^-\Phi_3^+)\cr
&+{1\over2}\xi_3(\Phi_{45}{\bar\Phi}_{45}+h_{45}{\bar h}_{45}
+D_{45}{\bar D}_{45}+\Phi_1^+{\bar\Phi}_1^++
\Phi_1^-{\bar\Phi}_1^-+\Phi_2^+{\bar\Phi}_2^++\Phi_2^-{\bar\Phi}_2^-
+\Phi_3^+{\bar\Phi}_3^+\cr
&+\Phi_3^-{\bar\Phi}_3^-)
+h_3{\bar h}_{45}\Phi_{45}+{\bar h}_3h_{45}{\bar\Phi}_{45}+\{{1\over 2}[\xi_1
(H_{19}H_{20}+H_{21}H_{22}+H_{23}H_{24}+H_{25}H_{26})\cr
&+\xi_2(H_{13}H_{14}+H_{15}H_{16}+H_{17}H_{18})]+\bar \Phi_{23}H_{24}H_{25}+
\Phi_{23}H_{23}H_{26}+h_2H_{16}H_{17}\cr
&+\bar h_2 H_{15}H_{18}+e^c_{L_2}H_8H_{29}+(V_1H_9+V_2H_{11})
H_{27}+V_6H_5H_{29}+\bar \Phi_{45}H_{17}H_{24}\cr
&+D_{45}H_{18}H_{21}+h_{45}H_{16}
H_{25}\} \hskip 9 cm (5)}$$
with a common normalization constant ${\sqrt 2}g$.

There are higher order (i.e. $N>3$) contributions to the superpotential which
can be calculated from the world--sheet correlators. Among other things, these
give masses to the quarks, Higgs doublets [\SLM,\NRT,\FM] and induce
quark mixing [\CKM] etc.

\bigskip
\centerline{\bf 3. SUSY}

In order to preserve SUSY at $M_P$, one has to satisfy a set of F and D
constraints.
The set of F and D constraints is given by the following equations:
$$\eqalignno{&D_A=\sum_k Q^A_k \vert \chi_k \vert^2={-g^2e^{\phi_D}
\over 192\pi^2}Tr(Q_A) {1\over{2\alpha^{\prime}}}&(6a) \cr
&D^{\prime j}=\sum_k Q^{\prime j}_k \vert \chi_k \vert^2=0 \qquad j=1
\ldots 5 &(6b) \cr
&D^j=\sum_k Q^j_k \vert \chi_k \vert^2=0 \qquad j=C,L,7,8 &(6c) \cr
&W={\partial W\over \partial \eta_i} =0 &(6d) \cr}$$
where $\chi_k$ and $\eta_i$ are the fields that do and do not get VEVs
respectively
and $Q^j_k$ are their charges. $2\alpha^{\prime}=g^2M_P^2/{8 \pi}=M^2$ and
$W$ is the superpotential. The charges
$Q^{j\prime}_k$ correspond to linear combinations of the original local
$U(1)_{r_j}$ which are non--anomalous whereas $Q_A$ corresponds to an
anomalous $U(1)$ local symmetry with $Tr(Q_A)=180$ [\MOD,\NRT]. From the DSW
mechanism, the D
constraint for the anomalous $U(1)_A$ gets an additional term proportional to
$Tr(Q_A)$. From Eq. (6a) we see that, $SO(10)$
singlet scalars must get VEVs $\sim g^2M/4 \pi \sim M/25$ in order to preserve
SUSY at $M_P$. The scale of the $SO(10)$ singlet VEVs is fixed by the $U(1)_A$
charges of the fields and the coefficient of the anomaly term. Since for a
generic
VEV, $\l \Phi \r \sim M/25$, we can expand the superpotential in the number of
terms (inverse powers of $M$) where order $N$ is suppressed by an order of
magnitude with respect to order $N-1$. This justifies our expansion in $N$.

The set of F constraints in the observable sector has been studied before
[\NRT]. One finds that SUSY requires (when either $\l H_{23} \r= \l H_{25}
\r=0$ or $\l H_{24} \r=\l H_{26} \r=0$ which as we will see below is
the case)
$$\l \Phi_{12} \r=\l \bar \Phi_{12} \r= \l \xi_3 \r=0 \eqno(7)$$
even though the number of fields is larger than the number of constraints.
Then, one is left with only three F constraints from the observable sector:
$$\eqalignno{&\bar \Phi_{23} \Phi_{13}+\bar \Phi_i^+ \bar \Phi_i^- =0 &(8a)\cr
            &\bar \Phi_{13} \Phi_{23}+ \Phi_i^+  \Phi_i^- =0 &(8b)\cr
            &\Phi_{45} \bar \Phi_{45}+ \Phi_i^+ \bar \Phi_i^+ +\Phi_i^-
 \bar \Phi_i^- =0 &(8c)} $$

In the hidden sector, on the other hand, we get the following F constraints:
$$\eqalignno{&H_{19}H_{20}+H_{23}H_{24}+H_{25}H_{26}=0 &(9a)\cr
            &H_{13}H_{14}+H_{17}H_{18}=0 &(9b)\cr
            &{1\over 2}\xi_1H_{24}+\Phi_{23}H_{26}=0 &(9c)\cr
            &{1\over 2}\xi_1H_{23}+\bar \Phi_{23}H_{25}=0 &(9d)\cr
            &{1\over 2}\xi_1H_{26}+ \bar \Phi_{23}H_{24}=0 &(9e)\cr
            &{1\over 2}\xi_1H_{25}+\Phi_{23}H_{23}=0 &(9f)\cr
            &{1\over 2}\xi_2H_{13}={1\over 2}\xi_2H_{14}={1\over 2}\xi_2H_{17}=
{1\over 2}\xi_2H_{18}=0  &(9g)} $$

As we will see in Section 5, the requirement of realistic (or non--zero) $b$,
$s$, $\mu$ and $\tau$ masses means that
$\xi_1$ and $\xi_2$ must get VEVs. Then, from the above F constraints we see
that $$\l H_{13} \r=\l H_{14} \r=\l H_{17} \r=\l H_{18} \r=0 \eqno(10)$$
in order to preserve SUSY at $M_P$. $H_{19}$ and $H_{20}$ are $5$ and $\bar
5$ of $SU(5)_H$ and obtain masses of $\l \xi_2 \r /2 \sim M$. Therefore, they
decouple from the spectrum before the hidden $SU(5)_H$ condenses at the scale
$\Lambda_H$ which means that $\l H_{19} H_{20} \r=0$.
For the rest, $H_{23},H_{24},H_{25},H_{26}$, from Eqs. (9c-f) we get the
following constraints:

a) When $\l \Phi_{23} \r \not =0$ and $\l \bar \Phi_{23} \r \not =0$, either
$\l H_{23} \r=\l H_{25} \r=0 $ or $\l H_{24} \r= \l H_{26} \r=0 $.

b) When one or both of $\Phi_{23},\bar \Phi_{23}$ have vanishing VEVs,
$\l H_{23} \r=\l H_{25} \r =\l H_{24} \r= \l H_{26} \r=0 $.

We stress that these SUSY constraints on the hidden sector VEVs are obtained
by requiring a realistic heavy quark and lepton spectrum (i.e. $\l \xi_{1,2}
\r \not =0$). Otherwise, Eqs. (9a-g) do not lead to useful SUSY constraints on
the hidden sector VEVs.

Since $\l H_{25} \r=0$ in order to have $h_{45}$ light as we will see in the
next section, $\l H_{23} \r=0$ in any case in this model.
$\l H_{24} \r$ and $\l H_{26} \r$ may or may not vanish depending on the VEVs
of $\Phi_{23}$ and $\bar \Phi_{23}$ as above.

\bigskip
\centerline{\bf 4. Higgs doublet masses}

The Higgs doublets in the model are $\{h_1,h_2,h_3,h_{45}\}$, their barred
counterparts (i.e. $\bar h_1$ etc.), $H_{15}$ and $H_{16}$. The light Higgs
doublets are determined by the Higgs doublet mass matrix  $h_i (M_h)_{ij}
\bar h_j$ where $h_i=(h_1,h_2,h_3,h_{45},H_{15})$ and $\bar h_j=(\bar h_1,
\bar h_2, \bar h_3,\bar h_{45},H_{16})$. From the cubic superpotential we get
$$M_h=\left(\matrix{&0 &0 &\bar \Phi_{13} &0 &0 \cr
                    &0 &0 &\bar \Phi_{23} &0 &0 \cr
                    &\Phi_{13} &\Phi_{23} &0 &\Phi_{45} &0 \cr
                    &0 &0 &\bar \Phi_{45} &0 &H_{25} \cr
                    &0 &0 &0 &0 &\xi_2 }\right) \eqno(11)$$
where the SUSY constraints Eq. (7) and Eq. (10) have been taken into account.
The entries in $M_h$ are the VEVs of the respective fields. The
matrix $M_h$ is diagonalized by a bi--unitary transformation $(SM_hT^{\dagger})
_{ij}=m_i \delta_{ij}$. The $h$ and $\bar h$ eigenstates and eigenvalues are
found by evaluating those of $M_hM_h^{\dagger}$ and $M_h^{\dagger}M_h$
respectively.

One sees that $H_{15}$ and $H_{16}$ will get masses (of $O(M)$) since $\l
\xi_2 \r \not=0$. In addition, $h_{45}$ will be heavy unless $\l H_{25}
\r=0$. $h_{45}$ must remain massless or light in order to get a
non--vanishing down quark mass matrix $M_d$. Therefore we impose $\l H_{25} \r
=0$ which we used in the last section. Then, generically, there are two
massless doublets:one linear combination
of $\{h_1,h_2,h_{45}\}$ and another linear combination of $\{\bar h_1,\bar h_2,
\bar h_{45}\}$. Note that $h_3$ ($\bar h_3$) gets mass and decouples unless
$\l \Phi_{13} \r=\l \Phi_{23} \r=\l \Phi_{45} \r=0$ ($\l \bar \Phi_{13} \r=
\l \bar \Phi_{23} \r=\l \bar \Phi_{45} \r=0$). As we will see in the next
section, non--vanishing quark mixing
imposes $\l \Phi_{45} \r \not=0$ so $h_3$ is necessarily heavy. In general,
one needs to give VEVs to any or all of $\{\bar \Phi_{13},\bar \Phi_{23},
\bar \Phi_{45}\}$ so that $\bar h_3$ is also heavy. The exact light doublet
states depend on the specific F and D flat direction chosen for the model. For
example if we take $\l \bar \Phi_{45} \r=
\l \Phi_{13} \r=0$, the light doublets are $h_{45},(-\l \bar \Phi_{23} \r/
\l \bar \Phi_{13} \r)h_1+h_2,\bar h_1$ and $\bar h_2+(\l \Phi_{23} \r/
\l \Phi_{45} \r)\bar h_{45}$. Different flat directions with two more vanishing
VEVs give different linear combinations which remain light. Thus, when two
more scalar VEVs (in addition to those that vanish due to SUSY constraints) are
zero, there
are two combinations of $h_i$ and $\bar h_i$ each which remain light.
One doublet of each kind must get mass at an intermediate scale lest there
is the danger of large contributions to flavor changing neutral currents from
Higgs exchanges.
Since this question is not directly related to the problem we examine, we
simply assume that this is the case and not expand on this point further.

Higgs doublets also get masses from higher order non--renormalizable terms.
Since only $\bar h_1$ or $\bar h_2$ and $h_{45}$ are relevant for quark mass
matrices we list only the terms involving them. There are terms which contain
hidden sector states with $U(1)_{Z^\prime}$ charge such as
$$h_2 \bar h_{45} \Phi_{45} H_{23}H_{26} \qquad \qquad \bar h_2 h_{45} \bar
\Phi_{45} H_{24}H_{25} \eqno(12a,b)$$
Similar terms for $h_1$ and $\bar h_1$ do not exist because of $U(1)_{r_1}$
conservation. The potentially dangerous terms (12a,b) vanish due to SUSY
constraints in the hidden sector.
In addition, there are terms which contain states from the sectors $b_j+2
\gamma+(I)$ such as
%\eqalignno{&h_1 \bar h_2 \Phi_{45}\Phi_1^+ T_1 \bar T_1 \qquad \qquad h_2
%\bar h_1 \Phi_{45}\bar \Phi_2^- T_2 \bar T_2 &(13a,b) \cr
$$h_1 \bar h_2 \Phi_{45}\bar \Phi_1^+ V_1 \bar V_1 \qquad \qquad h_2 \bar h_1
\Phi_{45}  \Phi_2^- V_2 \bar V_2 \eqno(13a,b)  $$
Also there are terms which induce a mixing i.e. $\mu$ terms such as
$$\eqalignno{&\bar h_1 h_{45} \bar \Phi_3^- T_3 \bar T_3 \qquad \qquad \bar
h_2 h_{45}  \Phi_3^+ T_3 \bar T_3 &(14a,b) \cr
&\bar h_1 h_{45}  \Phi_3^- V_3 \bar V_3 \qquad \qquad \bar h_2 h_{45}
\bar \Phi_3^+ V_3 \bar V_3 &(14c,d)}$$
In order to get enough quark mixing, one needs $\l V_j \r \sim 10^{16}~GeV$.
This means
that $\l V_j \r$ does not arise from condensation of $SU(3)_H$ (since
$\Lambda_H \sim 10^{10}~GeV$ for this model) but from the D constraints [\CKM].
The scale of scalar VEVs which arise from the D constraints is
$\sim M/25 \sim 10^{16}~GeV$ as fixed by the right hand side of Eq. (6a).
{}From Eqs. (13a,b) and higher order terms of the same kind we see that unless
we choose to give VEVs only to
one $V_j$ from each sector $j$, we get very large masses for the doublets which
couple to up--like quarks. In that case, they decouple from the low--energy
spectrum and up--like quarks cannot get masses. This is
phenomenologically unacceptable, so we give VEVs to only one $V_j$ from
each sector $b_j+2\gamma+(I)$. Thus, the dangerous terms in Eqs.
(13a,b) and (14c,d) vanish.

{}From Eq. (14a,b) we see that the Higgs mixing terms due to
$\l T_3 \bar T_3 \r$ give
an unacceptably large $\mu \sim 10^8~GeV$ for $\Lambda_H \sim 10^{14}~GeV$.
As a result, when $\bar h_1$ ($\bar h_2$) is light, we have to take
$\l \bar \Phi_3^- \r=0$ ($\l \Phi_3^+ \r=0$). Then, Higgs
mixing will arise from higher order terms which are smaller and acceptable.

\bigskip
\centerline{\bf 5. Light Quark Masses}

At the cubic order of the superpotential, there is only a mass term for the
heaviest up quark i.e. the top when only $\bar h_1$ or $\bar h_2$ remains
light.
The light Higgs doublet with $Q_L=1$ determines the heaviest generation.
At the quartic order there are no potential quark mass terms. At the
quintic order the following mass terms are obtained [\FM],
$$\eqalignno{&d_1Q_1h_{45}{\Phi}_1^+\xi_2
  {\hskip 2cm}d_2Q_2h_{45}{\bar\Phi}_2^-\xi_1 &(15a,b)\cr
             &u_1Q_1({\bar h}_{45}\Phi_{45}{\bar\Phi}_{13}+
  {\bar h}_2{\Phi}_i^+{\Phi}_i^-)&(15c)\cr
      &u_2Q_2({\bar h}_{45}\Phi_{45}{\bar\Phi}_{23}+
  {\bar h}_1{\bar\Phi}_i^+{\bar\Phi}_i^-)&(15d)\cr
      &(u_1Q_1h_1+u_2Q_2h_2)
  {{\partial W}\over{\partial\xi}_3}.&(15e)\cr}$$
The mass term from the cubic superpotential and the terms above give the mass
hierarchy between the two heavy quark generations. The heaviest up--like quark
mass term appears at $N=3$ whereas the mass terms for the other
quarks (except for the light ones) appear at $N=5$. The latter are suppressed
by
a factor of $\Phi^2/M^2 \sim (1/25)^2$ relative to the former and as a result
one can easily obtain the two or three orders of magnitude suppression
required.
The analysis of the non--renormalizable terms up to order $N=8$
shows that quark mixing terms are obtained for all generations [\CKM]. The
full list of terms that give mixing between the quarks have been given in
Ref. (\CKM) and will not be repeated here. In addition there are lepton mass
terms[\NRT]
$$e_1L_1 h_{45} \Phi_1^- \xi_2 \qquad \qquad e_2L_2 h_{45} \bar \Phi_2^+ \xi_1
\eqno(16a,b)$$

There are two possible scenarios with $\bar h_1$ or $\bar h_2$ as the light
doublet that couples to the up--like quarks. (An additional possibility is
that a linear combination of the two remains light. That case is easy to
examine once these two are understood.) In both cases the down--like
quarks couple to $h_{45}$ because of $Q_{\ell_j}$ quantum numbers.
In the first case, with $\bar h_1$ and $h_{45}$ as the light doublets, we get
with a suitable set of $SO(10)$ singlets with VEVs[\CKM]
%$\{\Phi_1^\pm, \bar
%\Phi_1^\pm, \Phi_2^-, \bar \Phi_2^\pm, \Phi_3^\pm, \bar \Phi_3^\pm, \Phi_{45},
%\bar \Phi_{13}, \Phi_{23}, V_1, \bar V_2, V_3, \xi_1, \xi_2\}$
(and the SUSY constraints from Eq. (7))

$$M_u\sim\left(\matrix{&\epsilon_1
&{{V_3{\bar V}_2\Phi_{45}\bar \Phi_3^+}\over{M^4}} &0\cr
&{{V_3{\bar V}_2\Phi_{45}\bar \Phi_2^+}\over{M^4}}
&{{{\bar\Phi}_i^-\bar \Phi_i^+}\over{M^2}}
&{V_1{\bar V}_2\Phi_{45}\bar \Phi_2^+}\over{M^4} \cr
&0 &{V_1{\bar V}_2\Phi_{45}{\bar\Phi}_1^+}\over{M^4}
&1\cr}\right)\l \bar h_1 \r \eqno(17)$$
and
$$M_d\sim\left(\matrix{&\epsilon_2
&{{V_3{\bar V}_2\Phi_{45}}\over{M^3}} &0\cr
&{{V_3{\bar V}_2\Phi_{45}\xi_1}\over{M^4}}
&{{{\bar\Phi}_2^-\xi_1}\over{M^2}} &{V_1{\bar V}_2\Phi_{45}\xi_i}\over{M^4} \cr
&0 &{V_1{\bar V}_2\Phi_{45}\xi_i}\over{M^4}
&{{\Phi_1^+\xi_2}\over{M^2}}\cr}\right)\l h_{45} \r \eqno(18)$$
where $\epsilon_{1,2}$ are small numbers to be determined below.
In the second case, where the light doublets are $\bar h_2$ and $h_{45}$, we
get
with the same set of VEVs[\CKM]
$$M_u\sim\left(\matrix{&\epsilon_1
&{{V_3{\bar V}_1\Phi_{45} \Phi_3^-}\over{M^4}} &0\cr
&{{V_3{\bar V}_1\Phi_{45} \Phi_1^-}\over{M^4}}
&{{{\Phi}_i^- \Phi_i^+}\over{M^2}}
&{V_2{\bar V}_1\Phi_{45} \Phi_1^-}\over{M^4} \cr
&0 &{V_2{\bar V}_1\Phi_{45}{\bar\Phi}_2^+}\over{M^4}
&1\cr}\right)\l \bar h_2 \r \eqno(19)$$
and
$$M_d\sim\left(\matrix{&\epsilon_2
&{{V_3{\bar V}_1\Phi_{45}}\over{M^3}} &0\cr
&{{V_3{\bar V}_1\Phi_{45}\xi_2}\over{M^4}}
&{{{\bar\Phi}_1^+\xi_2}\over{M^2}} &{V_2{\bar V}_1\Phi_{45}\xi_i}\over{M^4} \cr
&0 &{V_2{\bar V}_1\Phi_{45}\xi_i}\over{M^4}
&{{\bar \Phi_2^-\xi_1}\over{M^2}}\cr}\right)\l h_{45} \r \eqno(20)$$

As explained in Ref. (\CKM) the texture of $M_u$ and $M_d$ is a result of the
choice of $V_j, \bar V_j$s which get VEVs. (We remind that only one $V_j,\bar
V_j$ from each sector $b_j+2 \gamma$ can get a VEV due to constraints from
Higgs doublet masses.) Once the VEVs are chosen, there is
an effective discrete symmetry which protects the zeros in $M_u$ and $M_d$
except for light quark masses (i.e. $\epsilon_{1,2}$).
The up and down quark mass matrices are diagonalized
by bi--unitary transformations
$U_LM_uU_R^\dagger=D_u\equiv{\rm diag}(m_u,m_c,m_t)$ and
$D_LM_dD_R^\dagger=D_d\equiv{\rm diag}(m_d,m_s,m_b)$
with the CKM mixing matrix given by $V=U_LD_L^\dagger$.
Since we are only interested in order of magnitude results, we will loosely
take
$tan \beta =\l \bar h_{1,2} \r /\l h_{45} \r \sim 1$ in both cases.

{}From Eqs. (15a-e) and the mass matrices we see that in both cases the light
generation is fixed to be the one with index 3 since it is by far the lightest.
The heaviest generation is the one with index 1 (2) for the first (second)
case since the light Higgs doublet which couples to up--like quarks in the
cubic superpotential is $\bar h_1$ ($\bar h_2$).
If we consider only the observable sector of the spectrum, we find that there
are potential $Q_3u_3$ and $Q_3d_3$ terms such as
$$Q_3u_3 \bar h_1 \Phi_{13} \xi_3 \qquad \qquad Q_3d_3 h_{45}\bar \Phi_{13}
\Phi_3^+ \xi_3 \eqno(21a,b)$$
These terms potentially induce very large masses to up and down quarks. (For
generic VEVs, they give $m_u \sim GeV$ and $m_d \sim 100~MeV$.) Fortunately,
they vanish due to the SUSY constraints in the observable sector as
given by Eq. (7). The SUSY constraints gaurantee that all potential
$Q_3u_3$ or $Q_3d_3$ terms, arising only from the observable states, vanish to
all orders in $N$ as follows [\FM]. $Q_3,u_3$
and $d_3$ have $Q_{\ell_3}=1/2$ and the only $SO(10)$ singlets with
$Q_{\ell_3}$ charges in the observable sector are
$\Phi_{12},\bar \Phi_{12}$ and $\xi_3$ (since $h_3, \bar h_3$ get very large
masses and decouple at low energies). If these cannot get VEVs, then there
cannot be non--vanishing $Q_3u_3$ or $Q_3d_3$ terms to any order $N$ due to
conservation of $Q_{\ell_3}$. Actually
the fact that these terms vanish is welcome since if they did not they
would result in up and down quark masses which are too large.

Of course, there are other contributions once hidden sector states are taken
into account. For example, at $N=5$ and $N=6$, there are terms containing
hidden sector states with $U(1)_{Z^\prime}$ charge such as
$$Q_3u_3 \bar h_{45} H_{17}H_{24} \qquad \qquad Q_3d_3 h_{45}\Phi_3^+
H_{24}H_{25} \eqno(22a,b)$$
which potentially induce very large up and down masses (similar to those from
Eq. (21a,b)
in magnitude). These, too, vanish due to the SUSY constraints in the hidden
sector (with the additional requirement for a realistic heavy quark spectrum).
In both cases, when $H_{23},H_{25}$ or $H_{24},H_{26}$ have vanishing
VEVs, the terms in Eq. (22a,b) and all similar higher order terms containing
$H_i$ vanish.
Finally, we have terms which contain states from the hidden sectors
$b_j+2\gamma
+(I)$.  At $N=8$ we get terms with $T_j \bar T_j$ and $V_j \bar V_j$
$$\eqalignno{&Q_3u_3 \bar h_1 \Phi_3^+ \Phi_{45} \Phi_{13} T_3 \bar T_3 &(23a)
\cr
&Q_3u_3 \bar h_2 \Phi_3^- \Phi_{45} \Phi_{13} T_3 \bar T_3 &(23b) \cr
&Q_3d_3 h_{45} \bar \Phi_3^+ \Phi_3^- \Phi_{45} T_3 \bar T_3 &(23c) \cr
&Q_3u_3 \bar h_1  \Phi_3^- \Phi_{45} \Phi_{13} V_3 \bar V_3 &(23d) \cr
&Q_3u_3 \bar h_2 \Phi_3^+ \Phi_{45} \Phi_{13} V_3 \bar V_3 &(23e) \cr
&Q_3d_3 h_{45} \bar \Phi_3^- \Phi_3^- \Phi_{45} V_3 \bar V_3 &(23f) }$$
The terms containing $V_3 \bar V_3$ can potentially induce $m_u$ and $m_d$ of
$O(MeV)$. These vanish since only one $V_j$ or $\bar V_j$from every
sector can get a VEV as we saw before.
When $\bar h_2$ remains light, the term in Eq. (23a) vanishes due to the $\mu$
constraint, Eq. (9b), which gives $\l \Phi_3^+ \r=0$. The other two
terms are non--vanishing in general and are $\sim O(10^{-6}~MeV)$ for
$\Lambda_H \sim 10^{14}~GeV$. We see that light quark masses get
contributions either from the terms (23a-c) or from the off--diagonal terms in
$M_u$ and $M_d$ through diagonalization.

We stress that the potentially large
up and down mass terms in Eqs. (21) and (22) vanish because of the SUSY
constraints in the observable and hidden sectors. The set of vanishing VEVs due
to SUSY results in an effective
$Z_4$ symmetry under which $Q_3,u_3,d_3$ have charge $\alpha$ and $T_j,\bar T_j
,V_j,\bar V_j$ have charge $\alpha^3$ ($\alpha^4=1$). This discrete symmetry
eliminates the terms in Eqs. (21) and (22). Then, the only potential up and
down mass terms come either from Eqs. (23a-f) or from the mixing terms in $M_u$
and $M_d$.

We analyze the two cases seperately with the requirement
$sin \theta_c \sim 0.2$.
In the first case, when the light doublets are $\bar h_1$ and $h_{45}$, $M_u$
and $M_d$ are given by Eqs. (17) and (18) respectively. Quark mixing arises
mainly from $M_d$ and as a result we have
$$sin \theta_c \sim {{(M_d)_{12}}\over (M_d)_{22}} \qquad \hbox{and} \qquad
m_d \sim {{(M_d)_{12}(M_d)_{21}}\over (M_d)_{22}} \eqno(24)$$
Therefore, $m_d \sim 0.2 \times
(M_d)_{21}$. Since $(M_d)_{21} \sim (M_d)_{12} \l \xi_1 \r$ and
$(M_d)_{22}=m_s \sim 150~MeV$, we find that $m_d \sim 4 \times 10^{-5} \l \xi_1
\r \l h_{45} \r/M$. $\xi_1$ does not appear in F or D constraints and
therefore we can
choose $\l \xi_1 \r \sim M$ which gives $m_d \sim 4 ~MeV$, of the correct order
of magnitude. (From $m_{\mu} \sim 100~MeV$ we find that $\l \xi_1 \r \sim M$
requires $\l \bar \Phi_2^- \r \sim 10^{-3}M$ which is certainly possible but
an order of magnitude smaller than its natural value, $M/25$.)
One might think that the appearance of a large VEV, i.e. $\l \xi_1 \r \sim M$,
destroys the perturbative expansion in $N$. One can simply form terms at $N+1$
from terms at $N$ by adding $\xi_1$. This is not the case because string
(or world--sheet) selection rules require that, whenever one adds $\xi_1$ to
a non--vanishing string of fields, another field (with VEV $\sim M/25$)
accompany it. Then, one can form terms only at order $N+2$ (by adding $\xi_1$
to
an order $N$ term) and these are
suppressed relative to order $N$ terms by at least an order of magnitude.
Note that the contribution of the terms in Eq. (23a-c) is extremely
small ($\sim 10^{-6}~MeV$!) compared to the above.

The situation for $m_u$
is different. We find that $m_u \sim (M_u)_{12}(M_u)_{21}/(M_u)_{22}$ where
$$(M_u)_{12} \sim (M_d)_{12} {\l \Phi \r \over M} \sim 2
\times 10^{-4} \l h_{45} \r {\l \Phi \r \over M} \eqno(25)$$
On the other hand, since $\l \xi_1 \r \sim M$, from the $m_{\mu}$ term,
Eq.(16b), we obtain $\l \bar \Phi_2^+ \r \sim 10^{-3}M$.
Using $(M_u)_{22}=m_c \sim 1.5~GeV$ we get $m_u \sim 10^{-5}~MeV$ which is
six orders of magnitude smaller than the current up mass and too large to
solve the strong CP problem naturally. In order to solve the strong CP problem
one needs [\SCP]
$$\theta_{tot}{z\over {1+z}}<10^{-9} \eqno(26)$$
where $\theta_{tot}=\theta_{QCD}+\theta_{quarks}$ and $z=m_u/m_d$. For the
current quark masses $m_u \sim 5~MeV$ and $m_d
\sim 9~MeV$ one needs $\theta_{tot}<10^{-9}$ which is the strong CP problem.
It has been shown that there are no axions which can solve the strong CP
problem in standard--like superstring models [\AX]. An alternative solution is
$\theta_{tot}$ of $O(1)$ and $z < 10^{-9}$ which requires a very small or
vanishing up quark mass. We remind that such a small $m_u$ is compatible with
current algebra results [\GM]. Note
that since quark mixing comes from $M_d$, there is no constraint (lower bound)
on elements of $M_u$ from $sin \theta_c$.
It is the constraint $sin \theta \sim 0.2$ and the existence of $\l \xi_1 \r$
which can get a large ($\sim M$) VEV in $(M_d)_{21}$ that allows for an
acceptable $m_d$.

Another possibility is the case with $\l \bar \Phi_3^+ \r=0$.
$\l \bar \Phi_2^+ \r$ (or $(M_u)_{21}$) cannot vanish because
it appears in the $m_{\mu}$ term given by Eq. (16b). $\l \bar \Phi_3^+ \r$ (or
$(M_u)_{21}$) on the other hand, can
vanish since it does not appear in any term that must be non--zero. The only
constraints on $\l \bar \Phi_3^+ \r$ are the F and D constraints and these can
be satisfied whether $\l \bar \Phi_3^+ \r$ vanishes or not.
When $\l \bar \Phi_3^+ \r=0$ the leading contribution to $m_u$ comes from the
hidden sector condensate term in Eq. (23a). This gives
$m_u \sim 10^{-6}~MeV$ which is an order of magnitude smaller than before but
still not small enough to solve the strong CP problem naturally.
If, on the other hand, we choose a flat direction with $\l \Phi_3^\pm \r=
\l \bar \Phi_3^\pm \r=0$, then $m_u=0$ to up to $N=8$. In this case
there are no contributions to up mass either from $Q_3u_3$ terms or from
mixing in $M_u$ up to $N=8$. Does this mean that $m_u=0$ to all orders in $N$
for this
flat direction? The answer is no since there are higher order contributions to
up mass from $Q_3u_3$ terms (from $T_1 \bar T_1$ and $T_2 \bar T_2$
condensates) and mixing in $M_u$ from $N>8$. In fact there are $N=9$ terms
which
cannot vanish due to other phenomenological constraints such as heavy quark
and lepton masses etc. The non--vanishing $N=9$ terms give
$m_u \sim 10^{-7}~MeV$ which can be taken as a lower bound in this model.
Quite generally,
even if we choose a flat direction for which $m_u$ vanishes up to some order in
$N$, there will be non--vanishing contributions from higher orders. There are
no
flat directions for which $m_u$ vanishes to all $N$ if we demand realistic
heavy quark and lepton masses.

In the second case, where the light doublets are $\bar h_2$ and $h_{45}$, $M_u$
and $M_d$ are given by Eqs. (19)and (20) respectively. Again
$$sin \theta_c \sim {(M_d)_{12}\over (M_d)_{22}} \qquad \hbox{and} \qquad
m_d  \sim {{(M_d)_{21}(M_d)_{12}}\over (M_d)_{22}} \eqno(24)$$
but now the elements of $M_u$ and $M_d$ are different.
(Now generation with index 2 is the heaviest.) Requiring
$sin \theta_c \sim 0.2$ and taking $\l \xi_2 \r\sim M$, we find once again
$m_d \sim 4 \times 10^{-5} \l \xi_1 \r \l h_{45} \r/M$
which gives $m_d \sim 4~MeV$. Now, as in the first case, $m_u$ vanishes if
either $(M_u)_{12}$ or $(M_u)_{21}$ vanishes. $(M_u)_{21}$ is non--zero
because $\l \Phi_1^- \r$ appears in $m_{\mu}$. $(M_u)_{12}$, on the other hand,
can vanish since $\l \Phi_3^- \r$ does not have to get a VEV. The situation is
similar to the first case and using the same arguments we find that $m_u \sim
10^{-5}~MeV$ if $\l \Phi_3^- \r \not=0$. When $\l \Phi_3^- \r=0$, $m_u$ arises
mainly from the $Q_3u_3$ terms containing the hidden sector condensates which
give $m_u \sim 10^{-6}~MeV$ as before. The previous comments on solving the
strong CP problem and a vanishing $m_u$ to all orders for some flat direction
for the first case hold for this case too.

\bigskip
\centerline{\bf 6. Discussion and Conclusions}

We examined the light quark masses in a standard--like superstring model
in this paper. Our results are as follows. An acceptable $m_d$
(i.e. $\sim 5~MeV$) can be obtained for a family of flat
directions with $\l \xi_1 \r \sim M$ (or $\l \xi_2 \r \sim M$). $m_d$
arises from the off--diagonal terms in the down quark mass matrix. This is a
result of the fact that quark
mixing arises mainly from $M_d$ and $sin \theta_c \sim 0.2$. Then, $(M_d)_{12}$
is more or less fixed since $sin \theta_c \sim (M_d)_{12}/m_s$. In
addition $(M_d)_{21} =(M_d)_{12}\l \xi_2 \r$ where $\l \xi_2 \r$ is free since
it does not appear in F and D constraints. Thus, we can take $\l \xi_2 \r
\sim M$ and obtain an $MeV$ scale $m_d$.

The situation for $m_u$ is different. There are no constraints from quark
mixing on $(M_u)_{12}$ and $(M_u)_{21}$ so they are in general much smaller.
Moreover, $\l \xi_2 \r$ is replaced with $\l \bar \Phi_2^+ \r$ and
$\l \bar \Phi_3^+ \r$ (or with $\Phi_1^-$ and $\Phi_3^-$ in the second case)
which are about $M/10^3$ and $M/10$ respectively.
Consequently, from the off--diagonal terms in $M_u$, we find that, in either
case, $m_u \sim 10^{-5}~MeV$ which is six orders of magnitude smaller than
the current up mass. We stress that, in standard--like superstring models,
$m_u$
cannot be as large as its value from current algebra. Still, such a small up
mass is not in conflict with current algebra results. $m_u$ though very small,
is not small enough to solve the strong CP problem naturally.
In the first (second) case $m_u$ can be an order of magnitude smaller if
$\l \bar \Phi_3^+ \r=0$ ($\l \Phi_3^- \r=0$).
In that case, $m_u$ arises not from the off--diagonal terms in $M_u$ but from
$N=8$ terms with hidden sector condensates and $m_u<10^{-6}~MeV$.
We also find that for a F and D flat direction with $\l \Phi_3^\pm \r=\l
\bar \Phi_3^\pm \r=0$, $m_u=0$ up to $N=8$ but not for all $N$ since there are
always some non--zero $N=9$ terms which contribute to $m_u$. As a result, $m_u$
cannot vanish in this model. The $N=9$ terms give
$m_u \sim 10^{-7}~MeV$ which still is not small enough to solve the strong
CP problem naturally.

Why are the light quark masses so small compared to other quark masses or the
weak scale? We see that SUSY preservation in the observable and hidden sectors,
with the requirement of realistic heavy quark and lepton masses,
plays a very important role in this respect. In general, there are potentially
large mass terms for $u$ and $d$ from states in the observable sector. These
vanish to all orders due to the SUSY constraints in the observable sector.
There are additional terms coming from the hidden sector states with $U(1)_{Z^
\prime}$ charge. These, too, vanish due to SUSY constraints in the hidden
sector (using the requirement of a realistic heavy quark and lepton spectrum).
As a result, the only non--vanishing $Q_3u_3$ and $Q_3d_3$ terms
arise from VEVs of hidden sector condensates $\l T_j \bar T_j \r$ which break
SUSY dynamically in the hidden sector or from terms containing
$\l V_j \bar V_j \r$. The former are extremely small ($\sim 10^{-6}~MeV$)
since they are suppressed by $(\Lambda_H^2/M^2) \sim 10^{-8}$ with respect to
the others. The latter which can potentially give $MeV$ scale up and down
masses
vanish due to the constraints from an acceptable Higgs doublet spectrum.

There are also contributions to $m_u$ and $m_d$ from the off--diagonal terms in
up and down quark mass matrices (through diagonalization).
These off-diagonal elements or mixing terms arise from VEVs of hidden sector
states ($V_i \bar V_j$ etc.). In any case, the mixing terms are small compared
to the diagonal (or mass) terms i.e. $(M_{u,d})_{ij}<<(M_{u,d})_{ii}$
($1<i,j<3$) except for the $33$ terms, that is except for the light quark mass
terms.
As a result, the contribution of the mixing terms to the light quark masses is
much smaller than other quark masses.

Why is $m_u$ so much smaller than $m_d$ in this model? One reason is the
different contributions coming from the mixing terms in the mass matrices as
explained above. The other is the elimination of the terms containing
$\l V_j \bar V_j \r$ due to the constraints from an acceptable Higgs doublet
spectrum. These terms are the only ones which give an $MeV$ scale contribution
to $m_u$ in this model.

Our results can also be explained by using symmetry arguments.
SUSY constraints in the observable and hidden sectors force a number of VEVs
to vanish. This, in turn, results in an effective $Z_4$ symmetry
(with parameter $\alpha$, where $\alpha^4=1$). Under this $Z_4$, $Q_3,u_3,d_3$
have charge $\alpha$ and $T_j, \bar T_j, V_j, \bar V_j$ have charge $\alpha^3$.
Then, the only potential up and down mass terms come from terms which contain
$\l T_j \bar T_j \r$ and $\l V_j \bar V_j \r$ or from mixing terms as we saw.
Once different $ T_j$ and $ V_j $ get VEVs, the $Z_4$ symmetry is broken and
up and down quarks obtain masses.

We conclude that, in the standard--like superstring model considered, $m_u$
cannot be as large as a few $MeV$s. In fact it is at most $10^{-5}~MeV$. For
a family of flat directions or vacua $m_u$ can be as small as $10^{-7}~MeV$
but not much smaller. This range of up masses i.e. $10^{-7}~MeV<m_u<10^{-5}
{}~MeV$ cannot solve the strong CP problem naturally but is compatible with
current algebra results.
An acceptable $m_d$ at the $MeV$ scale can easily be obtained. We find that
the smallness of the light quark masses relative to the other quark masses is
connected to supersymmetry preservation in the observable and hidden sectors
and the requirement for a realistic haevy quark and lepton spectrum.

\bigskip
\centerline{\bf Acknowledgments}

This work is supported by a Feinberg Fellowship and the Department of
Physics. I thank Alon Faraggi for many useful discussions on this subject.
I also thank Yosi Nir for correcting a few points.
\vfill
\eject
\refout

\vfill
\eject

\end

\input tables.tex
\nopagenumbers
\magnification=1000
\baselineskip=18pt
\hbox
{\hfill
{\begintable
\ F \ \|\ SEC \ \|\ $SU(3)_C$ $\times$ $SU(2)_L$ \ \|\ $Q_C$ & $Q_L$ & $Q_1$ &
$Q_2$
 & $Q_3$ & $Q_4$  & $Q_5$ & $Q_6$
 \ \|\ $SU(5)$ $\times$ $SU(3)$ \ \|\ $Q_7$ &
$Q_8$  \crthick
$V_1$ \|\ ${b_1+2\gamma}+(I)$ \|(1,1)\|~0 & ~~0 & ~~0 & ~~${1\over 2}$ &
 ~~$1\over 2$ & ~~$1\over2$
 & ~~0 & ~~0 \|(1,3)\| $-{1\over 2}$ &
{}~~$5\over 2$   \nr
${\bar V}_1$ \|\                \|(1,1)\| ~~0 & ~~0 & ~~0 & ~~${1\over 2}$ &
 ~~$1\over 2$ & ~~$1\over2$
  & ~~0 & ~~0
  \|(1,$\bar 3$)\| ~~${1\over 2}$ &
$-{5\over 2}$  \nr
$T_1$ \|\                \|(1,1)\| ~~0 & ~~0 & ~~0 & ~~${1\over 2}$ &
 ~~$1\over 2$ & $-{1\over2}$
  & ~~0 & ~~0
  \|(5,1)\| $-{1\over 2}$ &
$-{3\over 2}$  \nr
${\bar T}_1$ \|\                \|(1,1)\| ~~0 & ~~0 & ~~0 & ~~${1\over 2}$ &
 ~~$1\over 2$ & $-{1\over2}$
   & ~~0 & ~~0
  \|($\bar 5$,1)\| ~~$1\over2$ &
{}~~${3\over 2}$  \cr
$V_{2}$ \|\ ${b_2+2\gamma}+(I)$ \|(1,1)\| ~~0 & ~~0 &
 ~~${1\over 2}$ & ~~0 &
 ~~$1\over 2$ & ~~0 &  ~~$1\over2$ & ~~0
 \|(1,3)\| $-{1\over 2}$ &
 ~~$5\over 2$  \nr
${\bar V}_{2}$ \|\                \|(1,1)\| ~~0 & ~~0 & ~~${1\over 2}$ & ~~0 &
 ~~$1\over 2$ & ~~0 & ~~$1\over2$ & ~~0
  \|(1,$\bar 3$)\| ~~${1\over 2}$ &
 $-{5\over 2}$  \nr
$T_{2}$ \|\                \|(1,1)\| ~~0 & ~~0 & ~~${1\over 2}$ & ~~0 &
 ~~$1\over 2$ & ~~0 & $-{1\over2}$ & ~~0
  \|(5,1)\| $-{1\over 2}$ &
 $-{3\over 2}$ \nr
${\bar T}_{2}$ \|\                \|(1,1)\| ~~0 & ~~0 & ~~${1\over 2}$ & ~~0 &
 ~~$1\over 2$ & ~~0 & $-{1\over2}$ & ~~0
  \|($\bar 5$,1)\| ~~$1\over 2$ &
 ~~${3\over 2}$  \cr
$V_{3}$ \|\ ${b_3+2\gamma}+(I)$ \|(1,1)\| ~~0 & ~~0 & ~~${1\over 2}$ &
 ~~${1\over 2}$ & ~~0 & ~~0 & ~~0 & ~~${1\over2}$
       \|(1,3)\| $-{1\over 2}$ &
 ~~$5\over 2$  \nr
${\bar V}_{3}$ \|\                \|(1,1)\| ~~0 & ~~0 & ~~${1\over 2}$ &
 ~~${1\over 2}$ & ~~0 & ~~0 & ~~0 & ~~$1\over2$
 \|(1,$\bar 3$)\| ~~${1\over 2}$
 & $-{5\over 2}$  \nr
$T_{3}$ \|\                \|(1,1)\| ~~0 & ~~0 & ~~${1\over 2}$ &
 ~~${1\over 2}$ & ~~0 & ~~0 & ~~0 & $-{1\over2}$
       \|(5,1)\|
 $-{1\over 2}$ & $-{3\over 2}$  \nr
${\bar T}_{3}$ \|\                \|(1,1)\| ~~0 & ~~0 & ~~${1\over2}$ &
 ~~${1\over 2}$ & ~~0 & ~~0 & ~~0 & $-{1\over2}$
   \|($\bar 5$,1)\| ~~$1\over 2$
 & ~~${3\over 2}$
 \endtable}
\hfill}
\bigskip
\parindent=0pt
\hangindent=39pt\hangafter=1
{\it Table 1.} Massless states from the sectors $b_j+2\gamma$,
and their quantum numbers.

\vfill
\eject

\end
\bye

\input tables.tex
\nopagenumbers
\magnification=1000
\baselineskip=18pt
\hbox
{\hfill
{\begintable
\ F \ \|\ SEC \ \|\ $SU(3)_C$ $\times$ $SU(2)_L$ \ \|\ $Q_C$ & $Q_L$ & $Q_1$ &
$Q_2$
 & $Q_3$ & $Q_4$  & $Q_5$ & $Q_6$
 \ \|\ $SU(5)$ $\times$ $SU(3)$ \ \|\ $Q_7$ &
$Q_8$  \crthick
$H_{13}$ \|\ $b_1+b_3+\alpha$  \|(1,1)\| $-{3\over 4}$ &
{}~~$1\over2$
& $-{1\over4}$ & ~~${1\over 4}$ & $-{1\over 4}$ & ~~0
 & ~~0 & ~~0 \|(1,3)\| ~~${3\over 4}$ & ~~$5\over 4$   \nr
$H_{14}$ \|\ $\pm \gamma+(I)$  \|(1,1)\| ~~$3\over4$ & $-{1\over2}$ & ~
$1\over4$
& $-{1\over 4}$ & ~~$1\over 4$ & ~~0 & ~~0 & ~~0
  \|(1,$\bar 3$)\| $-{3\over 4}$ & $-{5\over 4}$  \nr
$H_{15}$ \|\                \|(1,2)\| $-{3\over4}$ & $-{1\over2}$ &
$-{1\over4}$
& ~~${1\over 4}$ & $-{1\over 4}$ & ~~0 & ~~0 & ~~0
  \|(1,1)\| $-{1\over 4}$ & $-{15\over 4}$  \nr
$H_{16}$ \|\                \|(1,2)\| ~~$3\over4$ & ~~$1\over2$ & ~~$1\over4$
& $-{1\over 4}$ & ~~$1\over 4$ & ~~0 & ~~0 & ~~0
  \|($1,1)\| ~~$1\over4$ & ~~${15\over 4}$  \nr
$H_{17}$ \|\                  \|(1,1)\| $-{3\over4}$ & ~~$1\over2$ &
 $-{1\over 4}$ & $-{3\over4}$ & $-{1\over 4}$ & ~~0 & ~0& ~~0
 \|(1,1)\| $-{1\over 4}$ & $-{15\over 4}$  \nr
$H_{18}$ \|\                \|(1,1)\| ~~$3\over4$ & $-{1\over2}$ & ~
${1\over 4}$
& ~~$3\over4$ & ~~$1\over 4$ & ~~0 & ~~0 & ~~0
  \|(1,1$)\| ~~${1\over 4}$ & ~~${15\over 4}$  \cr
$H_{19}$ \|\ $b_2+b_3+\alpha$  \|(1,1)\| $-{3\over4}$
& ~~$1\over2$
& ~~${1\over 4}$ & $-{1\over4}$ & $-{1\over 4}$ & ~~0 & ~0& ~~0
  \|(5,1)\| $-{1\over 4}$ & ~~${9\over 4}$ \nr
$H_{20}$ \|\ $\pm \gamma+(I) $    \|(1,1)\| ~~$3\over4$ & $-{1\over2}$ &
$-{1\over 4}$
& ~~$1\over4$ & ~~$1\over 4$ & ~~0 & ~0& ~~0 \|($\bar 5$,1)\| ~~$1\over 4$ &
 $-{9\over 4}$  \nr
$H_{21}$ \|\                 \|(3,1)\| ~~$1\over4$ & ~~$1\over2$ & ~
${1\over 4}$
& $-{1\over 4}$ & $-{1\over4}$ & ~~0 & ~~0 & ~~0
       \|(1,1)\| $-{1\over 4}$ &
 $-{15\over 4}$  \nr
$H_{22}$ \|\                \|($\bar 3$,1)\| $-{1\over4}$ & $-{1\over2}$
& $-{1\over 4}$ & ~~${1\over 4}$ & ~~$1\over4$ & ~~0 & ~~0 & ~~0
 \|(1,1)\| ~~${1\over 4}$ & ~~${15\over 4}$  \nr
$H_{23}$ \|\                \|(1,1)\| $-{3\over4}$ & ~~$1\over2$ & ~
${1\over 4}$
& $-{1\over 4}$ & ~~$3\over4$ & ~~0 & ~~0 & ~~0       \|(1,1)\|
 ~~${1\over 4}$ & ~~${15\over 4}$  \nr
$H_{24}$\|\                \|(1,1)\| ~~$3\over4$ & $-{1\over2}$ & $-{1\over4}$
& ~~${1\over 4}$ & $-{3\over4}$ & ~~0 & ~~0 & ~~0 \|(1,1)\| ~~$1\over 4$
 & ~~${15\over 4}$ \nr
$H_{25}$ \|\                \|(1,1)\| $-{3\over4}$ & ~~$ 1\over2$ &~~$1\over4$
&
{}~~$3\over4$ & $-{1\over4}$ & ~~0 & ~~0 & ~~0 \|(1,1)\| $-{1\over4}$ &
$-{15\over4}$ \nr
$H_{26}$ \|\                \|(1,1)\| ~~$3\over4$ &  $-{1\over2}$ &
$-{1\over4}$ &  $-{3\over4}$ & ~~$1\over4$ & ~~0 & ~~0 & ~~0 \|(1,1)\|
{}~~$1\over4$ & ~~$15\over4$
 \endtable}
\hfill}
\bigskip
\parindent=0pt
\hangindent=39pt\hangafter=1
{\it Table 2.} States with $Z^{\prime}$ charge and their
quantum numbers.

\vfill
\eject

\end
\bye